\documentclass[11pt]{article}

\usepackage{amsmath}
\usepackage{amsfonts}
\usepackage{amssymb}

\def\be{\begin{equation}}
\def\ee{\end{equation}}

\usepackage{paper2e}

\begin{document}
\begin{titlepage}

\preprint{HUTP-03/A079}

\title{Ghost Inflation}

\author{Nima Arkani-Hamed$^{\rm a}$, Paolo Creminelli$^{\rm a}$, Shinji Mukohyama$^{\rm
a}$ and Matias Zaldarriaga$^{\rm a,b}$}

\address{$^{\rm a}$Jefferson Laboratory of Physics, Harvard University\\
Cambridge, Massachusetts 02138}

\address{$^{\rm b}$ Center for Astrophysics, Harvard University \\
Cambridge, Massachusetts 02138}

\begin{abstract}
We propose a new scenario for early cosmology, where an
inflationary de Sitter phase is obtained with a ghost condensate.
The transition to radiation dominance is triggered by the ghost
itself, without any slow-roll potential. Density perturbations are
generated by fluctuations around the ghost condensate and  can be
reliably computed in the effective field theory. The fluctuations
are scale invariant as a consequence of the de Sitter symmetries,
however, the size of the perturbations are parametrically
different from conventional slow-roll inflation, and the inflation
happens at far lower energy scales. The model makes definite
predictions that distinguish it from standard inflation, and can
be sharply excluded or confirmed by experiments in the near
future. The tilt in the scalar spectrum is predicted to vanish
($n_s=1$), and the gravity wave signal is negligible. The
non-Gaussianities in the spectrum are predicted to be observable:
the 3-point function is determined up to an overall ${\cal O}(1)$
constant, and its magnitude is much bigger than in conventional
inflation, with an equivalent $f_{\rm NL} \simeq 100$, not far from
the present WMAP bounds.
\end{abstract}

\end{titlepage}

\section{Introduction}

Inflation is an attractive paradigm for early cosmology. An early
de Sitter phase of the universe elegantly solves the horizon and
flatness problems. Further imagining that this phase is driven by
a slowly varying scalar field with a flat potential gives a
beautiful mechanism for generating the observed density
perturbations, which have their origin in the quantum fluctuations
of this light field during inflation. This minimal picture of
slow-roll inflation predicts the fluctuations to be nearly scale
invariant and Gaussian, in excellent agreement with observation.

While this qualitative picture is very attractive, it is
frustratingly difficult to make much sharper predictions that
could definitively exclude or confirm it in future experiments.
There are nevertheless some generic expectations for what should
be seen.  For instance, in most models, the tilt in the scalar
spectrum is non-vanishing, and is typically of order $n_s - 1 \sim
(1/N_{e}) \sim 2 \times 10^{-2}$, where $N_e$ is the number of e-folds
to the end of inflation when typical cosmological scales leave the horizon. 
The current bounds on $n_s - 1$ are getting close to this size, while the 
Planck satellite will improve the sensitivity to $\sim 3 \cdot 10^{-3}$. 
Slow-roll inflation leads to the expectation that Planck should find $n_s \neq 1$, but this is
not a hard prediction, and if Planck does not find a deviation
from $n_s = 1$ we can not definitely exclude it. If a tilt is
found, it would be interesting to look for finer structures in the
power spectrum, such as bumps or wiggles of various kinds.
However, one can always engineer bumps and wiggles in the inflaton
potential to reproduce these features in the power spectrum, so
there is nothing here either that can exclude or confirm the
standard picture with certainty.  The observation of a
gravitational wave signal in the CMBR would be extremely exciting
and it would indicate that the energy density driving the accelerated expansion
is very high, of order $M_{\rm GUT}^4$.
Minimal slow roll inflation predicts a relationship between the
tilt of gravity wave spectrum $n_g - 1$ and the scalar one $n_s - 1$,
and if this is found it would be a confirmation of the standard
picture. However, we would have to be lucky for the gravity wave
spectrum to be observable in near future experiments, to say
nothing of the {\it tilt} in the gravity wave spectrum, and if the
inflationary prediction is found {\it not} to hold, it would by no
means exclude the theory; deviations from this relationship can
come about by integrating out new physics between the Planck and
Hubble scales during inflation \cite{Kaloper:2002uj,Shiu:2002kg}.

The minimal picture of slow-roll inflation {\it does} make one
very sharp prediction, however: the spectrum is predicted to be
Gaussian to a very high degree. For single field inflation models,
the 3-point function was recently calculated to leading order in
the slow-roll parameters by Maldacena \cite{Maldacena:2002vr}. 
The dimensionless size of the effect was found to be parametrically at least of the same
order as $\delta \rho/\rho$ itself, (and in fact further
suppressed by slow-roll parameters). The prediction is very firm,
since the leading non-Gaussianities are generated by the
interaction of the inflaton with gravity, and are therefore
determined by the symmetries of the theory. However, the magnitude
is about a factor of $\sim 100$ smaller than what might be
observable in the future experiments. Because of this firm
prediction, however, non-Gaussianities are the best and most
exciting place to look for deviations from the standard picture of
slow-roll inflation. Indeed there are by now a number of different
ways of generating density perturbations from sources other than
the fluctuations of the slowly rolling inflaton, and in these
models it is possible to get larger (and observable)
non-Gaussianities \cite{Bernardeau:2002jy,Lyth:2002my,Dvali:2003em}. Any observation 
of non-Gaussianities necessarily implies the presence of additional light fields and a
non-standard mechanism for generating density perturbations.

Given the difficulty of conclusively verifying the standard
inflationary paradigm, it is worth looking for alternatives,
especially if they are more predictive in some way. Given the
ease with which a de Sitter phase of expansion obliterates the 
horizon and flatness problems, it would be nice to
keep this, but is there an alternative to the standard slow-roll
picture?

In this paper we will describe such a model. Recently, a new possibility
of having de Sitter phases in the universe in a way differing from
a cosmological constant has been proposed \cite{ghost}, which also modifies gravity
in the infrared in a non-trivial and consistent way.  It can be
thought of as arising from a derivatively coupled ghost scalar field $\phi$ 
which ``condenses" in a background where it has  a non-zero velocity
\begin{equation}
\langle \dot{\phi} \rangle = M^2 \, \, \to \langle \phi \rangle =
M^2 t \;.
\end{equation}
The novelty here is that, unlike other scalar fields, the velocity
$\dot{\phi}$ does {\it not} redshift to zero as the universe
expands, it stays constant, and indeed the energy momentum tensor
is identical to that of a cosmological constant. However, the
ghost condensate is {\it not} a cosmological constant, it is a
physical fluid with a physical fluctuations $\pi$ defined as
\begin{equation}
\phi = M^2 t + \pi \;.
\end{equation}
The ghost condensate then gives an alternative way of realizing
de Sitter phases in the universe. The symmetries of the theory
allow us to construct a systematic and reliable effective
Lagrangian for $\pi$ and gravity at energies lower than the ghost
cut-off $M$. Neglecting the interactions with gravity, the
effective Lagrangian for $\pi$ (around flat space) has the form
\begin{equation}
\label{eq:piaction}
S = \int \!d^4x \;\frac12 \dot{\pi}^2 - \frac{\alpha^2}{2 M^2}
(\nabla^2 \pi)^2 - \frac{\beta}{2 M^2} \dot{\pi} (\nabla \pi)^2 + \cdots
\end{equation}
where $\alpha$ and $\beta$ are order one coefficients.
Note that this Lagrangian is non-Lorentz invariant, as should be
expected, since the background $\dot{\phi} = M^2$ breaks Lorentz
invariance spontaneously (the $\pi$ field can be thought as the
Goldstone boson for this symmetry breaking). The low-energy
dispersion relation for $\pi$ is of the form
\begin{equation}
\omega^2 = \alpha^2 \frac{k^4}{M^2} \;.
\end{equation}
It is straightforward to couple this healthy sector to gravity,
which leads to a variety of interesting modifications of gravity
in the IR, including antigravity and oscillatory modulation of the
Newtonian potential at late times and large distances
\cite{ghost}.

Given that this theory provides an alternative way of having a
de Sitter phase, it is natural to try to use the field $\phi$ as the
inflaton. Since we have a different value for the Hubble scale
during inflation than the present, the shift symmetry on $\phi$
must be broken. However, it is technically natural to assume that
there is a potential $V(\phi)$ which has the form {\em e.g.} $V(\phi) =
V_0$ for $\phi < 0$, $V(\phi) = 0$ for $\phi > 0$. The shift
symmetry is still good for $\phi < 0$ and $\phi > 0$, so that this 
shape is radiatively stable. Of course
this is an exaggeration of a smooth shape for the potential, and
we could have variations on each side over scales smaller than the
cutoff $M$ of the ghost theory. The theory inflates for $\phi <
0$, and evolves to radiation dominance for $\phi > 0$. Note
that we need not assume that $V_0$ itself is of order $M^4$. For
instance, as in hybrid inflation, $\phi$ could couple to matter
with shift-symmetry breaking couplings around $\phi=0$; these
couplings can trigger a phase transition at $\phi=0$ and $V_0$ 
can be the energy of the false vacuum in the matter sector. We will not
concern ourselves with describing any specific model for the
transition or with justifying the different energy scales, $M$ and $V_0$, 
of the model; it suffices to say that the set-up is completely
technically natural, {\em i.e.} radiatively stable.  From this minimal set-up, 
however, we can make definite and experimentally relevant predictions for the
density perturbations, which is our main focus in this
paper.

Let us begin with a qualitative picture and estimates for the size
and level of non-Gaussianity in the perturbations, and later
derive all the results more formally. There are two important
differences here from ordinary inflation. First, there is no
slow-roll. Even in the approximation where the potential for
$\phi<0$ is {\it exactly flat}, $\dot{\phi} = M^2$ is non-zero,
and $\phi$ is continuously heading towards $\phi=0$ where inflation
ends, continuing with about the same velocity afterwards. As
we will see in a moment, density perturbations are generated here
from the fluctuations of the $\pi$ field, and are again scale
invariant as in inflation. However, since there is no need for any
slope in the potential, there is no reason to expect any tilt in
the scalar spectrum. 

The second important difference with standard inflation concerns the size of
the fluctuations in $\phi$ (or equivalently $\pi$). Since the
effective Lagrangian for $\pi$ is non-relativistic, and in
particular there are no $k^2$ spatial kinetic terms, the
fluctuations of $\pi$ are less suppressed than in a relativistic
theory. In a relativistic theory, the fact that scalar fields have
scaling dimension 1 tells us that the size of the fluctuations of a
scalar field inside a region of size $R$ is given by $\sim 1/R$,
similarly at a frequency $E$ it is given by $E$. In usual
inflation, the inflaton fluctuations freeze when they have a typical 
energy $E \sim H$, so that their typical size is $\delta\phi \sim H$. We can 
determine the size of the fluctuations in our case by a simple scaling argument familiar
from power-counting for non-relativistic effective theories \cite{Polchinski:1992ed}.
Suppose we scale energies by a factor of $s$, $E \to s E$, or
alternatively $t \to s^{-1} t$. Clearly, because of the $\omega^2
\propto k^4$ dispersion relation, we have to scale $k$
differently, $k \to s^{1/2} k$ or $x \to s^{-1/2} x$. We then
determine the scaling dimension of $\pi$ by requiring the
quadratic action to be invariant, and we find that $\pi$ has
scaling dimension $1/4$
\begin{equation}
\pi \to s^{1/4} \pi \;.
\end{equation}
Now, $\pi$ has {\it mass} dimension $1$, so the fluctuations at
frequencies of order the cutoff $M$ is $\delta \pi_M \sim M$.
But the fact that $\pi$ has scaling dimension $1/4$ tells us that
the fluctuation at a lower energy $E$ is $\delta \pi_E  \sim (E
M^3)^{1/4}$. In particular, the size of ghost fluctuations that 
freeze, as usual, by Hubble friction when its frequency in of order $E \sim H$, is
\begin{equation}
\delta \pi_H \sim (H M^3)^{1/4} \;.
\end{equation}
Of course, for consistency of the effective theory, we must have $H
\ll M$; note that as expected these fluctuations are much larger
than $H$ in this limit.

Let us now estimate the size of the density perturbations. As
usual, $\phi$ fluctuates and is stretched out until the Hubble
damping becomes important at frequency $E \sim H$; note that this
does {\it not} correspond to $k \sim H$ but rather, from the
dispersion relation, $k \sim \sqrt{H M}$. The fluctuation $\delta \pi_H$ 
causes inflation to end at slightly different times in different places, and 
so as usual we have the estimate
\begin{equation} 
\frac{\delta \rho}{\rho} \sim H \delta t =
\frac{H \delta \pi_H}{\dot{\phi}} \;. 
\end{equation} 
Now both of the differences with the usual story come into play. First,
$\dot{\phi} = M^2$, having nothing to do with slow-roll
parameters. Second, $\delta \pi_H$ is much larger. We then find
\begin{equation} 
\label{eq:drho}
\frac{\delta \rho}{\rho} \sim
\left(\frac{H}{M}\right)^{5/4} \;,
\end{equation} 
which can be compared with the usual inflationary case
\begin{equation} 
\frac{\delta \rho}{\rho} \sim \frac{H}{M_{\rm Pl} \sqrt{\epsilon}} \;, 
\end{equation} 
where $\epsilon$ is the usual slow-roll parameter: $\epsilon \equiv 
M_{\rm Pl}^2/2 \cdot (V'/ V)^2$. 
Since $\phi$ continues to have the same velocity today, leading to the IR modifications 
of gravity described in \cite{ghost}, $M$ is limited to be at most 10 MeV or so:
we must have $H \sim 1$ keV at largest for $\delta \rho/\rho \sim
10^{-5}$.  We therefore generate the correct magnitude density
perturbations at far lower energy scales than usual inflation. Indeed,
the maximum energy scale is of order $V_0 \sim H^2 M_{\rm Pl}^2 \sim
(1000 \;{\rm TeV})^4$. This implies that the gravity wave signal is
completely negligible.

We can then reproduce the spectrum of perturbations, with the
sharp predictions that (a) $|n_s - 1| = 0$ and (b) there are no
gravitational waves. The observation of any of these would
decisively rule out our model.

Things get more interesting when we look at the non-Gaussianities
in the model. Because all the scales are so much smaller than the
Planck scale, there is no effect coming from gravitational
interactions. The dominant effect comes from the trilinear
interaction in the $\pi$ effective Lagrangian
\begin{equation} 
\frac{\beta}{M^2} \dot{\pi} (\nabla \pi)^2 \;.
\end{equation}
This leads to a non-zero three-point function for the 
density perturbations. To get an idea of the dimensionless size
of this effect, we need to find the dimensionless size of this
coupling at an energy of order $H$. We can do this easily since we
know the scaling dimension of $t,x$ and $\pi$: we find that
the coefficient of this operator has scaling dimension $1/4$;
it is an irrelevant operator, but just barely! The dimensionless
size of this interaction at the cutoff $M$ is $\sim 1$; scaling it
down to an energy of order $H$ we get an estimate for the non-gaussianity
of the perturbations
\begin{equation} 
{\rm NG} \sim \left(\frac{H}{M}\right)^{1/4} \sim \left(\frac{\delta
\rho}{\rho}\right)^{1/5} \;.
\end{equation} 
These are much larger than in standard inflation, where ${\rm NG} \sim \epsilon \cdot
(\delta\rho/\rho) $! In fact, this estimate suggests that the non-gaussianity may be {\it too} 
large, $\sim 10^{-1}$, but in fact the complete analysis we do below shows that it is
naturally just at the level of the present WMAP constraints. The estimate clearly shows the physical
origin of the effect, as being due to the unusual fluctuations and
scaling dimensions of interactions in the $\pi$ effective theory.
Furthermore, this operator is the dominant (least irrelevant)
operator at low energies. As such, the dependence of the
three-point function on momenta can be unambiguously predicted up
to small corrections, as we will see in detail below.

Before turning to the detailed analysis of our scenario, we point out that the idea of 
an accelerated expansion of the universe driven by a non-minimal kinetic term was 
already discussed under the name of k-inflation \cite{Armendariz-Picon:1999rj}. 
K-inflation is based on some assumed form for the higher derivative terms, which can 
only be justified by the knowledge of the fundamental theory. On the contrary 
our model is based on a consistent effective field theory around the ghost condensate, 
so that all the predictions are {\em not} UV sensitive.

\section{Generation of density perturbations}
Quantum fluctuations of the $\pi$ field in de Sitter space become,
as usual, classical fluctuations at large scales, when they freeze
and become constant. In ref.~\cite{ghost} it was shown that the
gravitational potential in longitudinal gauge, $\Phi = \Psi$, decays
to zero outside the horizon. This means that $\pi$ fluctuations do
not gravitate at superhorizon scales in the pure de Sitter
background. This is quite intuitive: in the limit in which the
shift symmetry $\pi \rightarrow \pi + {\rm const.}$ is exact,
different parts of the Universe with a different value of $\pi$
are equivalent, so that the metric is not perturbed.

This is rather different from what happens in standard inflation, where the
surfaces of constant inflaton are also of constant energy density
and the gravitational potential, therefore, does not decay to zero
outside the horizon. Our model is similar to the usual slow-roll
inflation in the limit in which the potential becomes completely
flat, keeping the produced density perturbations constant:
$\epsilon \rightarrow 0$ with $H/(M_{\rm Pl} \sqrt{\epsilon})= {\rm
const}$. In this limit the comoving surfaces of constant inflaton
have the only meaning of setting the time to the end of inflation,
but there is no metric perturbation before reheating: $\Phi
\rightarrow 0$ as $\epsilon \rightarrow 0$.

The calculation of density perturbations will therefore proceed as
usual: we can calculate the gauge invariant quantity $\zeta$
\cite{Bardeen:qw}, proportional to the spatial curvature of
comoving surfaces through the position dependent time shift 
\begin{equation}
\label{eq:shift} 
\zeta = - \frac{H}{\dot\phi} \pi \;. 
\end{equation} 
This quantity will remain constant outside the horizon, independently
of the details of the reheating process and it will seed all the
perturbations we observe today. For a simple check of the conservation of $\zeta$ 
in the limit of instantaneous reheating, see the Appendix.

The situation is in some sense simpler than in conventional
slow-roll inflation. In that case we are forced to switch to the
$\zeta$ variable at horizon crossing, because we know that $\zeta$
is constant outside the horizon, while the inflaton
perturbation will keep on evolving because of the non-zero
potential. The evolution of the inflaton perturbation outside the
horizon is such that it cancels the time evolution of $H/\dot\phi$
in (\ref{eq:shift}) to give a constant $\zeta$. In our case both
$\zeta$ and $\pi$ are constant outside the horizon before the end
of inflation as $H$ and $\dot\phi$ are time-independent. So we can
also easily follow the surface of constant $\pi$ until reheating.

Let us now calculate the spectrum of density perturbations. In presence of the 
ghost condensate gravity is modified in the IR and this modification is characterized
by a typical time scale $\Gamma^{-1} $, with $\Gamma \sim M^3/M_{\rm Pl}^2$, and a typical 
length scale $m^{-1}$, with $m \sim M^2/M_{\rm Pl}$ \cite{ghost}. The two scales are 
different because the condensate obviously breaks Lorentz invariance. To avoid large
modification of gravity nowadays we have to require $\Gamma < H_0$, where $H_0$ is the present
Hubble constant \cite{ghost}.  Obviously this implies that gravity is not modified during inflation, {\em i.e.}
\begin{equation}
\Gamma \ll m \ll H \;.
\end{equation}
This relation is equivalent to the decoupling limit $M_{\rm Pl} \rightarrow \infty$, 
keeping $H$ fixed. Therefore we
can study the $\pi$ Lagrangian alone, without considering the
backreaction on the metric through the Einstein equations. At
linear level we get 
\begin{equation} 
\label{eq:pi} 
\ddot\pi_k +3 H \dot\pi_k + \frac{\alpha^2}{M^2} \left(\frac{k^4}{a^4}\right) \pi_k = 0 \;,
\end{equation} 
where $k$ is the comoving wavevector and $a$ the scale factor. This equation describes
a free field in de Sitter space, with the modified dispersion relation $\omega^2 \propto k^4$.

To calculate the spectrum of $\pi$ fluctuations, we quantize the
field as usual 
\begin{equation} 
\label{eq:quantize} \pi_k(t) = w_k(t) \hat a_k
+ w_k^*(t) \hat a_{-k}^\dagger \;, 
\end{equation} 
where $w_k(t)$ satisfies (\ref{eq:pi}). A qualitative study of eq.~(\ref{eq:pi}) tells us
that the wavefunction goes to a constant when the expansion rate
$H$ exceeds the oscillation frequency $\alpha M^{-1} (k/a)^2$:
due to the modified dispersion relation the $\pi$ field freezes
when its physical wavelength is of order $(\alpha^{-1} H
M)^{-1/2}$ and not $H^{-1}$ as usual.

We can easily estimate the asymptotic amplitude of $w_k$, which
gives the spectrum of the $\pi$ perturbations. Well before
freezing eq.~(\ref{eq:pi}) describes an harmonic oscillator with
slowly varying frequency and with a friction term. Both the
variation of frequency and the friction reduce the oscillator
energy. Using the equipartition of kinetic and potential energy,
it is easy to obtain the time variation of the total energy $E$ of
the oscillator 
\begin{equation} 
\label{eq:redshift} 
\dot E(t) = -5 H E(t) \;,
\end{equation} 
which implies that the quantity $E(t) \cdot a^5$ is constant
before freezing. We can rewrite this using the amplitude of
oscillation $\delta\pi(t)$ and the frequency $\omega(t)$ as 
\begin{equation}
\label{eq:const} 
\delta\pi(t)^2 \omega(t)^2 a^5 = {\rm const} \;. 
\end{equation} 
Now we can get the asymptotic amplitude using this conservation law between 
one time well before freezing (where we set $a=1$) and the freezing point $\omega(t) 
\sim H$. At the first point we use the usual flat-space normalization of the wavefunction $\delta\pi
\sim \omega^{-1/2}$ and we get the final result 
\begin{equation}
\label{eq:power} 
\delta \pi_H (k) \sim \frac{(H M^3\alpha^{-3})^{1/4}}{k^{3/2}} \;, 
\end{equation}
compatible with what we got in the introduction using the scaling dimension
of the operators.  The spectrum of $\pi$ fluctuations is scale invariant.
This is not a surprise; it is just the consequence of time translation invariance 
in the de Sitter background; every mode freezes at a fixed amplitude when it reaches a certain 
physical size and this is enough to give a scale invariant spectrum, independently of the dispersion 
relation of the $\pi$ field and the fact that its modes freeze at a scale smaller than the
de Sitter horizon. 

The qualitative behavior above can be checked by the explicit solution of
eq.~(\ref{eq:pi}). To simplify the equation we can go to conformal
time $d\eta = dt/a$ (we take $\eta = - (a H)^{-1}$ during
inflation) and write the equation for the variable $u_k \equiv w_k
\cdot a$. We get 
\begin{equation} 
\label{eq:confeq} 
u''_k + \left(\frac{\alpha^2 H^2 k^4}{M^2} \cdot \eta^2 -
\frac2{\eta^2}\right) u_k =0 \;, 
\end{equation} 
where the derivatives are now taken with respect to $\eta$. The solution of
eq.~(\ref{eq:confeq}) with the correct flat space limit for
very short wavelength ({\em i.e.} $\eta \rightarrow -\infty$) is
given by 
\begin{equation} 
\label{eq:explicit} 
u_k(\eta) = \sqrt{\frac\pi8} \sqrt{-\eta} \; H^{(1)}_{3/4}\left(\frac{H k^2 \alpha}{2 M}
\eta^2\right) \;, 
\end{equation} 
where $H$ is the Hankel function, linear combination of Bessel functions: 
$H^{(1)}_\nu(x) = J_\nu(x)+ i Y_\nu(x)$. The spectrum of $\pi$ can be calculated 
from the asymptotic behavior of (\ref{eq:explicit}) in the limit $\eta
\rightarrow 0$. The Bessel function $J_{-3/4}$ is dominant and we
get 
\begin{equation} 
\label{eq:limit} 
u_k(\eta \rightarrow 0) \simeq \frac{i\sqrt{2 \pi}}{\Gamma(1/4)} \left(\frac{M}{\alpha
H}\right)^{3/4} k^{-3/2} \cdot \eta^{-1} \;. 
\end{equation} 
The $\pi$ spectrum is therefore given by 
\begin{equation} 
\label{eq:pispok} 
P_\pi = a^{-2} \frac{k^3}{2 \pi^2} |u_k(\eta \rightarrow 0)|^2 = \frac{(H
M^3 \alpha^{-3})^{1/2}}{\pi \, \Gamma^2(1/4)} \;. 
\end{equation}  
From this we get the primordial curvature spectrum through (\ref{eq:shift}) 
\begin{equation}
\label{eq:curspok} 
P_{\cal{R}}^{1/2} = \frac1{\sqrt{\pi}
\Gamma(1/4)} \frac{(H^5 M^3 \alpha^{-3})^{1/4}}{\dot\phi} =
\frac1{\sqrt{\pi} \Gamma(1/4)} \left(\frac{H}{M}\right)^{5/4}
\alpha^{-3/4} \;. 
\end{equation} 
The COBE normalization fixes $P_{\cal{R}}^{1/2} \simeq 4.8 \cdot 10^{-5}$.

\section{Non-gaussianities}
One of the sharpest prediction of inflation, at least if the
inflaton itself is responsible for the generation of density
perturbations, is that these density fluctuations should have
a probability distribution very close to gaussian.
There are two reasons behind this property. First of all the inflaton field must have a
very shallow potential which allows a slow-roll phase; this means
that the inflaton self-interactions are very small so that it
behaves like a free field. Non-linearities coming from General
Relativity, even if suppressed by the Planck scale, are in fact
bigger than those coming from the inflaton potential, but anyway
too small to be experimentally testable: the fluctuations are 
gaussian up to a level of $10^{-6}$ \cite{Maldacena:2002vr}. 
One can imagine that
additional self-interactions for the inflaton are coming from
higher dimension operators with derivatives: these operators are
compatible with slow-roll if they respect a shift symmetry on the
inflaton $\phi \rightarrow \phi + {\rm const}$. The contribution
of these operators cannot be very big anyway: the lowest dimension
operator, $(\nabla\phi)^4$, has scaling dimension 2, so that its
contribution is suppressed by $(H/\Lambda)^2$, where $\Lambda$ is the cut-off
scale. As $\Lambda$ cannot be too small if we want the classical motion
of the inflaton to be under control ({\em i.e.} $\dot\phi \ll
\Lambda^2$), we get a contribution which cannot be much bigger than
$10^{-5}$ \cite{Creminelli:2003iq}.

In our model we do not have any potential for $\phi$ during
inflation and the only non-linearities come from higher
dimension operators. As discussed in the introduction, these
interactions are actually much more relevant than in standard
inflation, because of the non-relativistic scaling of the
operators. The most relevant (or better least irrelevant)
operator is 
\begin{equation} 
\label{eq:cubic} 
\frac{\beta}{M^2} \dot\pi (\nabla\pi)^2  \;. 
\end{equation} 
It has scaling dimension $[{\rm energy}]^{1/4}$, all the others have
dimension bigger or equal to $1/2$ so that they will be less
important in calculating the non-gaussianities of the density
perturbations. The very small dimension of this operator tells
us that this will be much more important than higher dimension
operators in standard inflation, which have at least dimension
2. Again we implicitly assumed a decoupling limit, neglecting
the backreaction of $\pi$ on the metric: in this way we neglect
all the non-linearities intrinsic to General Relativity, which are
suppressed by $M_{\rm Pl}$ and therefore subleading with respect to the
$\pi$ self-interactions. 

Let us estimate the size of the non-gaussianities comparing the non-linear 
correction $\dot\pi (\nabla\pi)^2$ with the linear terms in
eq.~(\ref{eq:pi}), close to the freezing point $\omega \simeq H$. The corrections 
will be given by the ratio 
\begin{equation}
\label{eq:nlest} 
\frac{\beta M^{-2} (H P_\pi^{1/2})(M H \alpha^{-1} P_\pi)}{H^2 P_\pi} \simeq
\frac\beta{\sqrt{\pi}\Gamma(1/4)}\left(\frac{H}{M}\right)^{1/4}
\alpha^{-7/4} \simeq 3 \cdot 10^{-2} \cdot \beta \cdot \alpha^{-8/5} \;, 
\end{equation}
where in the last equality we used the COBE normalization
eq.~(\ref{eq:curspok}). Parametrically we obtain the same
estimate we got in the introduction using the scaling dimension of
the operator. Taking $\alpha \sim \beta \sim 1$ the level of
non-gaussianity is quite substantial and close to what is allowed
by experiments. 

Despite the order one uncertainty we expect that the non-gaussian
contributions are experimentally detectable. The angular
distribution of the 3-point function clearly depends on the
wavefunction of the $\pi$ field and on the specific trilinear
interaction and could in principle be used as a smoking gun of
the model. We therefore turn to the explicit calculation of
the 3-point function given by the interaction $\dot\pi (\nabla\pi)^2$.

Before doing the explicit calculation we must clarify the relation
between $\pi$ and the gauge invariant quantity $\zeta$. At the end
we are obviously interested in the 3-point function of $\zeta$, as
this quantity remains constant outside the horizon (also at
non-linear order \cite{Salopek:1990jq}) and it is directly related
to the fluctuations we observe today. In doing the calculations it
is much easier to calculate the 3-point function of the $\pi$
field and only at the end turn to the $\zeta$ variable. Physically
we calculate the 3-point function of the $\pi$ field after its
fluctuations are frozen outside the horizon then, before the end
of inflation, we switch to comoving coordinates where $\pi$ is
constant on surfaces of fixed time, as $\pi$ fixes when inflation
ends at each point. The position dependent time shift between the
two sets of coordinates is given by (\ref{eq:shift}). It is very
important to note that this relation is valid also at non-linear
order in $\pi$ as $\dot\phi$ and $H$ are constant so that they do
not depend on $\pi$. This is different from what happens in
slow-roll inflation where the relation between $\zeta$ and the
inflaton fluctuations is non-linear \cite{Maldacena:2002vr}. In
our case there are no additional non-linearities going from the
variable $\pi$ to $\zeta$: the two 3-point functions are just
proportional through the relation (\ref{eq:shift}).

In the calculation of the 3-point function we start from the
interaction Lagrangian 
\begin{equation} 
\label{eq:intL} 
{\cal{L}}_{\rm int} =
-\beta\frac{e^{Ht}}{2 M^2} \left[\dot\pi(\nabla\pi)^2\right] \;. 
\end{equation}
We are interested in the value of the $\langle\pi\pi\pi\rangle$ correlator at a time
$t$ when the fluctuations are frozen outside the horizon, so we
have to evolve it starting from an initial time deep inside the
horizon 
\begin{equation} 
\label{eq:evol}
\langle\pi_{k_1}(t)\pi_{k_2}(t)\pi_{k_3}(t) \rangle = -i
\int_{t_0}^t dt' \left\langle
\left[\pi_{k_1}(t)\pi_{k_2}(t)\pi_{k_3}(t) , \int d^3x \;
{\cal{H}}_{\rm int}(t') \right]\right\rangle \;. 
\end{equation} 
The expression can be evaluated using the decomposition
(\ref{eq:quantize}) of the $\pi$ field, giving
\begin{align}
\label{eq:integral} 
\langle\pi_{k_1}\pi_{k_2}\pi_{k_3} \rangle & =
\frac{i \beta}{M^2} (2 \pi)^3 \delta^3 \big(\sum_i \vec k_i \big)
w_{k_1}(0) w_{k_2}(0) w_{k_3}(0)  \\ \nonumber & \cdot
\int_{-\infty}^0 d\eta \frac1{H\eta} w_{k_1}^*(\eta)
w_{k_2}^*(\eta) w^{\prime *}_{k_3}(\eta) (\vec k_1 \cdot \vec k_2)
+ {\rm symm.} + c.c.
\end{align}
The sum over the two analogous expressions with the derivative
acting on the other wavefunctions is indicated with '+symm'. The
integral must evaluated with the prescription that the oscillating
functions become exponentially decreasing, {\em i.e.} below the
negative real axis: this fixes the vacuum of the interacting
theory at $\eta = -\infty$. Using the explicit expression for the
wavefunctions (\ref{eq:explicit}) we get to the final expression
\begin{align}
\label{eq:NGfinal} \langle\pi_{k_1}\pi_{k_2}\pi_{k_3} \rangle & =
-\frac{2 \sqrt{2} \pi^{3/2}}{\Gamma(1/4)^3} \frac{H^5 \beta}{M^2}
\left(\frac{M}{\alpha H}\right)^4 (2\pi)^3\delta^3 \big(\sum_i
\vec k_i \big) \\ \nonumber & \frac1{\prod_i k_i^3}
\int_{-\infty}^0 d\eta \;\eta^{-1} F^*(\eta)
F^*\left(\frac{k_2}{k_1} \eta\right) F^{\prime
*}\left(\frac{k_3}{k_1}\eta\right) k_3 (\vec k_1 \cdot \vec k_2) +
{\rm symm.} + c.c.
\end{align}
where \begin{equation} \label{eq:F} F(x) = \sqrt{\frac{\pi}{8}} (-x)^{3/2}
H^{(1)}_{3/4}(x^2/2) \;. 
\end{equation} 
We can finally translate the result into the $\zeta$ variable, using 
eq.~(\ref{eq:shift}) 
\begin{align}
\label{eq:zetafinal} \langle\zeta_{k_1}\zeta_{k_2}\zeta_{k_3}
\rangle & = \frac{2 \sqrt{2} \pi^{3/2} \beta}{\Gamma(1/4)^3}
\left(\frac{H}{\alpha M}\right)^4 (2\pi)^3\delta^3 \big(\sum_i
\vec k_i \big) \\ \nonumber & \frac1{\prod_i k_i^3} 2 \,{\rm Re}\,
\int_{-\infty}^0 d\eta \;\eta^{-1} F^*(\eta)
F^*\left(\frac{k_2}{k_1} \eta\right) F^{\prime
*}\left(\frac{k_3}{k_1}\eta\right) k_3 (\vec k_1 \cdot \vec k_2) +
{\rm symm.}
\end{align}

As there are no poles one can rotate the contour of integration
along the direction $\propto (-1-i)$ so that it converges
exponentially.

In the analysis of the data (see {\em e.g.} \cite{Komatsu:2003fd})
it is usually assumed that the non-gaussianities come from a field
redefinition 
\begin{equation} 
\label{eq:fNL} 
\zeta(x) = \zeta_g(x) - \frac35 f_{\rm
NL} (\zeta_g^2(x) -\langle\zeta_g^2\rangle) \;, 
\end{equation} 
where $\zeta_g$ is gaussian. This pattern of non-gaussianity, which is local in
real space, is characteristic of models in which the
non-linearities develop outside the horizon. 
This happens for all the models in which the fluctuations of an additional
light field, different from the inflaton, contribute to the curvature
perturbations we observe. In this case non-linearities come from the
evolution of this field outside the horizon and from the conversion mechanism
which transforms the fluctuations of this field into density perturbations.
Both these sources of non-linearity give a non-gaussianity of the form
(\ref{eq:fNL}) because they occur outside the horizon. Examples of this 
general scenario are the curvaton models \cite{Lyth:2002my}, models with
fluctuations in the reheating efficiency \cite{Dvali:2003em} and multi-field
inflationary models \cite{Bernardeau:2002jy}.
In the data analyses (\ref{eq:fNL}) is taken as a simple ansatz and limits are therefore imposed on
the scalar variable $f_{\rm NL}$. The angular dependence of the 
3-point function in momentum space implied by (\ref{eq:fNL}) is given by
\begin{equation} 
\label{eq:fNLFour} \langle\zeta_{k_1}\zeta_{k_2}\zeta_{k_3}
\rangle = (2\pi)^3\delta^3 \big(\sum_i \vec k_i \big) \; (2\pi)^4
\, \big(-\frac35 f_{\rm NL} P_{\cal{R}}^2\big) \;\frac{4 \sum_i
k_i^3}{\prod_i 2 k_i^3} \;. 
\end{equation} 
In our case the angular distribution is much more complicated than in the previous
expression so the comparison is not straightforward.

We can nevertheless compare the two distributions
(\ref{eq:zetafinal}) and (\ref{eq:fNLFour}) for an equilateral
configuration and define in this way  an ``effective'' $f_{\rm
NL}$ for $k_1 = k_2 = k_3$. Evaluating numerically the integral in
(\ref{eq:zetafinal}) for $k_1 = k_2 = k_3$ and using COBE
normalization $P_{\cal{R}}^{1/2} \simeq 4.8 \times 10^{-5}$ and
eq.~(\ref{eq:curspok}) we get to 
\begin{equation}
\label{eq:eqfNL} 
f_{\rm NL}^{\rm equil.} \simeq 85 \cdot \beta \cdot \alpha^{-8/5} \;. 
\end{equation} 
The result is rather suppressed, taking $\alpha \sim \beta \sim 1$ with 
respect to the estimate (\ref{eq:nlest}), which would give $f_{\rm NL} \simeq 10^3$. 
There are two reasons tor this suppression: first of all $\dot\pi
\rightarrow 0$ near freezing and this suppresses the integral in
(\ref{eq:zetafinal}). Second the wavefunction is slightly
suppressed with respect to its natural size for $\eta \rightarrow
0$ and this gives an additional small suppression. The result must
be considered just as an estimate of the effect because order one
factors are not under control. On the other hand the angular
dependence in (\ref{eq:zetafinal}) is well defined and different from all the models generating 
observable non-gaussianities through an additional light field; the
angular pattern represents a potential smoking gun of the model.

The present limit on the non-gaussianity parameter from the WMAP
collaboration \cite{Komatsu:2003fd} 
\begin{equation} 
\label{eq:fNLlimit} 
-58 < f_{\rm NL} < 138 \quad {\rm at \;95\% \;C.L.} 
\end{equation} 
is quite close to the estimate (\ref{eq:eqfNL}).

We expect that the limit would change appreciably using for
the analysis the angular pattern (\ref{eq:zetafinal}). The reason
is that our pattern is qualitatively quite distinct from the one parametrized by
$f_{\rm NL}$. The behavior in the limit in which one of the wavevector goes to zero
is, for example, rather different from what happens in
(\ref{eq:fNLFour}). In this limit one of the fluctuations has a very
long wavelength, it exits the horizon and freezes much before the
other two and acts as a sort of background. Let us take $k_3$ very
small and assume that a spatial derivative acts on this background
in the interaction Lagrangian (\ref{eq:intL}). The 2-point
function $\langle\pi_1 \pi_2\rangle$ depends on the position on
the background wave and it is proportional to $\partial_i \pi_3$
at linear order. This variation of the 2-point function along the
$\pi_3$ wave is averaged to zero in calculating the 3-point
function $\langle\pi_{k_1}\pi_{k_2}\pi_{k_3}\rangle$, because the
spatial average $\langle\pi_3\partial_i\pi_3\rangle$ vanishes. So
we are forced to go to second order and we therefore expect the
integral in (\ref{eq:zetafinal}) to go as $k_3^2$, making the full
3-point function proportional to $k_3^{-1}$, after we multiply for
the $\pi_3$ spectrum. Note that taking the time derivative in
(\ref{eq:intL}) to act on the background gives a subleading
contribution as the wavefunction goes to a constant as $k_3^3$. On
the other hand the ``local'' behavior (\ref{eq:fNLFour}) goes as
$k_3^{-3}$ in the limit we have considered. In our model the
3-point function is generated by a derivative interaction which
favors the correlation of modes freezing roughly at the same time,
while the correlation is suppressed for modes of very different
wavelength. The same happens in standard inflation when we study
the non-gaussianity generated by higher derivative interactions
\cite{Creminelli:2003iq}. The implication of very different
angular patterns for the experimental limits on non-gaussianities
deserves further studies \cite{future}.

Before closing this section we want to comment on an assumption we 
have implicitly made so far. We have imposed on the effective Lagrangian 
for $\pi$, eq.~(\ref{eq:piaction}), a discrete symmetry $\pi \rightarrow -
\pi$, $t \rightarrow -t$. Without this assumption we should consider
an additional trilinear operator with the same scaling dimension as 
$\dot\pi (\nabla\pi)^2$:
\begin{equation}
\frac{\gamma}{M^3}(\nabla\pi)^2 \nabla^2 \pi \;,
\end{equation}
where $\gamma$ is an additional order one coefficient.
It is straightforward to repeat the analysis in this case; the qualitative
behavior of the 3-point function remains the same.

The $\pi \rightarrow - \pi$, $t \rightarrow -t$ symmetry is equivalent to 
$\phi \rightarrow -\phi$ for the ghost action. In absence of this symmetry
we can write CPT violating operators in the effective Lagrangian \cite{ghost}.
This leads us to a possible interesting effect in the CMBR which is independent 
of how density perturbations are generated: the rotation of the polarization vector 
of photons traveling from the surface of last scattering to us. The origin of this 
effect is that the ghost background $\langle\dot\phi\rangle \neq 0$ breaks the
CPT symmetry \cite{ghost} besides the Lorentz one, so that a term
$\epsilon^{0\mu\nu\rho} F_{\mu\nu} A_\rho$ for the photon
field is allowed. This term gives a different dispersion relation for the
two circular polarizations and it thus rotates the axis of a linearly polarized photon. 
Some limits on the size of this operator come from the observation of distant quasars and radio
galaxies \cite{Wardle:1997gu}: the rotation must be smaller than roughly one degree in one
Hubble time. Better limits should come in the future
from the CMBR experiments observing polarization \cite{Lue:1998mq}.
Temperature fluctuations on the last scattering surface always give
a certain degree of polarization to the radiation, with a typical
``gradient'' pattern called E-mode. The rotation of
the polarization transforms E-modes into ``curl'' or B-modes giving a
correlation between temperature fluctuations
and B-modes which would otherwise be zero. While the present sensitivity
is comparable to the limits from distant
quasars and radio galaxies, planned experiments could reach a sensitivity
of $10^{-3}$ degrees in an Hubble time \cite{Lue:1998mq}.
We stress that the size of the CPT violating operator cannot be predicted
as it depends on the coupling of the ghost sector
to ordinary matter.

\section{\label{sec:tilt}Tilting the potential}
So far we have assumed that the inflationary potential for $\phi$
is {\it exactly} flat, {\em i.e.} that the shift symmetry for $\phi$ is
not broken except at $\phi =0$. As we have emphasized,
unlike slow-roll inflation, no tilt in the potential is needed
either to end inflation or to generate acceptable density
perturbations. However, given that the shift symmetry on $\phi$
must be broken in any case, there may well be a tilt in the
potential for $\phi < 0$, although it is technically natural for
this tilt to be as small as we like. It is thus interesting to see
how our previous results are modified in the presence of a
non-zero slope. We expect, in the limit of very big tilt, to recover  
the predictions of standard slow-roll inflation, where the classical motion
of $\phi$ is dominated by the potential term. As a first approximation, we will assume that the
potential has a constant slope $V^\prime =$ const. The equation of
motion for the homogeneous perturbations $\pi$ then becomes
\begin{equation}
\ddot{\pi} + 3 H \dot{\pi} + V^\prime = 0
\end{equation}
and we quickly reach the solution for which
\begin{equation}
\dot{\pi} = - \frac{V^\prime}{3 H} \;.
\end{equation}
We see that, because of the potential, we have changed the
velocity of the field $\phi$; for $V^\prime < 0$, we have
increased the velocity\footnote{We will not study the other possibility, $V^\prime >0$, 
which leads to a wrong sign spatial kinetic term for $\pi$ \cite{ghost}.}. Note that, in order to stay within the
regime of validity of the effective theory, the velocity of $\pi$ should 
still be small compared with $M^2$; this motivates us to define the small
parameter
\begin{equation}
\delta^2 \equiv \frac{-V^\prime}{3 H M^2} \;.
\end{equation}
Now, with this non-zero value of $\dot{\pi}$ in the background,
the dispersion relation for $\pi$ is modified by the cubic
interaction 
\begin{equation}
\langle \dot{\pi} \rangle \frac{(\nabla \pi)^2}{M^2} \;,
\end{equation}
so we have (we take $\alpha = \beta = 1$)
\begin{equation}
\omega^2 = \frac{k^4}{M^2} + \delta^2 k^2 \;.
\end{equation}
As long as the $k^4$ term dominates down to frequencies
$\omega \sim H$, the $\delta$ term can be justifiably neglected;
this happens as long as
\begin{equation}
\delta \lesssim (H/M)^{1/2} \;.
\end{equation}
For larger $\delta$ the new $k^2$ term dominates before the
modes freeze out, at a cross-over frequency $\omega_{\rm cross}$
determined by
\begin{equation}
\omega_{\rm cross} \simeq \delta^2 M \;.
\end{equation}
We can estimate the size of the perturbations in this case easily,
as it is now just like the usual case for standard inflation, with the only
difference that the spatial momenta are rescaled as $k \to \delta \cdot
k$. Thus,
\begin{equation}
\delta \pi_H \sim \frac{H}{\delta^{3/2}} \;,
\end{equation}
which reduces to the usual result of standard inflation $\delta\pi_H \sim H$
for $\delta \rightarrow 1$, as we expected.
For the density perturbations we get
\begin{equation}
\label{eq:tiltpert}
\frac{\delta \rho}{\rho} \sim \frac{H \delta \pi_H}{\dot{\phi}}
\sim \frac{H^2}{M^2 \delta^{3/2}} \;.
\end{equation}
So, ranging over all values of $\delta$, $\delta \rho/\rho$ goes
from $(H/M)^{5/4}$ (see eq.~(\ref{eq:drho})) to $(H/M)^2$.

Let us now estimate the corrections to the spectral index. It is straightforward to check that corrections
coming from the variation of $H$ are completely negligible: the variation of the potential in one 
Hubble time is of order $\Delta V \sim \delta^2 M^4$, which is very small with respect to the 
total vacuum energy $V_0$. Bigger corrections can come from the variation of $\dot\phi$ in 
eq.~(\ref{eq:tiltpert}). The effect is of order $V^{\prime\prime}/H^2$ and limits on 
$V^{\prime\prime}$ come again from the validity of the effective field theory description
\begin{equation}
\delta^2 M^2 H \simeq |V^\prime| \gtrsim |V^{\prime\prime}| \cdot \Delta\phi = |V^{\prime\prime}| 
\cdot (M^2/H) N_e  \quad \Rightarrow \quad |V^{\prime\prime}| \lesssim \delta^2 H^2/N_e \;,
\end{equation}
where $N_e$ is the typical number of e-folds.
So that deviations from the flat spectrum can be as big as $\delta^2/N_e$: 
\begin{equation} 
|n_s-1| \lesssim \delta^2/N_e \;.
\end{equation}
The biggest contribution to the tilt comes however from the variation of $\delta$. Assuming
a variation of $\delta$ during observable inflation of order of $\delta$ itself, we get a contribution
to the tilt of order 
\begin{equation} 
|n_s-1| \lesssim 1/N_e \;.
\end{equation}
Therefore, in the regime $\delta \gtrsim (H/M)^{1/2}$, a rather big deviation from scale invariance 
is possible.

We can finally also estimate the dimensionless size of the
non-Gaussian effects, which still come from the same cubic
interaction (\ref{eq:cubic}) as before. For the frequencies $\omega_{\rm cross} <
\omega < M$ where the $k^4$ quadratic spatial kinetic term
dominates, this operator still has scaling dimension $1/4$, but
for $H < \omega < \omega_{\rm cross}$, the $k^2$ quadratic term
dominates and the operator has its familiar scaling dimension of
$2$. Therefore, the dimensionless size of the non-Gaussianity in
this case is
\begin{equation}
\label{eq:NGtilted}
{\rm NG} \sim \left(\frac{\omega_{\rm cross}}{M}\right)^{1/4} \times
\left(\frac{H}{\omega_{\rm cross}} \right)^2 \sim \frac{H^2}{M^2
\delta^{7/2}} \sim \frac{\delta \rho}{\rho} \times
\frac{1}{\delta^2}
\end{equation}
so we see that the non-Gaussianities are always parametrically
enhanced relative to $\delta \rho/\rho$. This is to be contrasted
with slow-roll inflation, where the leading non-Gaussianity is in
fact parametrically suppressed relative to $\delta \rho/\rho$ by
slow-roll factors. The same result eq.~(\ref{eq:NGtilted}) can be obtained directly 
evaluating the importance of the non-linear interaction $\dot\pi (\nabla \pi)^2$
with respect to free field terms at horizon crossing.

We note that the result for $\delta \rightarrow 1$ is what we get in
conventional slow-roll inflation, with the addition of higher derivative terms suppressed
by a cut-off scale $M \simeq \dot\phi^{1/2}$ \cite{Creminelli:2003iq}.
This limit can in fact be approached starting from a conventional slow-roll model with the 
addition of higher-dimension operators compatible with the shift symmetry of the inflaton. These 
additional operators will be suppressed by a typical scale $M$ (note that now we do not have
$\dot\phi \sim M^2$, because the motion of the inflaton is {\em not} dominated by these higher 
derivative terms). The slow-roll picture makes sense for $M^2 \gg \dot\phi$, otherwise all the 
operators become relevant for the classical motion of the inflaton. In the limit $M^2 \sim \dot\phi 
\simeq V^\prime/H$ we lose control of the theory, but we qualitatively approach our ghost model for 
$\delta \simeq 1$; higher dimension operators become as important as the tilt in the potential to describe 
the classical motion of the inflaton. 

In general if we consider higher derivative interactions suppressed by a typical scale $M$ and a potential
tilt $V^\prime$, we can look at our model and conventional slow-roll inflation as the
two extreme limits $M^2 \simeq \dot\phi \gg V^\prime/H$ (the potential tilt is irrelevant) and 
$M^2 \gg \dot\phi \simeq V^\prime/H$ (higher derivative terms are irrelevant). In both the limits we have a trustworthy effective field
theory description and they merge for $V^\prime/H \sim M^2$, {\em i.e.} $\delta \sim 1$, when the potential
tilt and the higher derivative terms give a comparable contribution to the inflaton motion. In this
regime both descriptions break down. In the table we summarize 
the predictions for the tilt in the scalar spectrum and the level of non-gaussianity 
going from the ghost inflation limit to the usual slow-roll case.

\begin{table}
\begin{center}
\begin{tabular}{|c|c|c|c|c|}
\hline  & $\delta < (H/M)^{1/2}$ & $(H/M)^{1/2} < \delta < 1$ & $\delta \sim 1$ & Standard slow-roll \\ \hline \hline
$|n_s-1|$ & 0 & $1/N_e$ & $1/N_e$ & $1/N_e$ \\  NG & $10^{-2}$ & $10^{-5} \delta^{-2}$ & $10^{-5}$ & $10^{-5} \epsilon$ \\
\hline 
\end{tabular}
\end{center}
\caption{Order of magnitude predictions for the tilt in the scalar spectrum and the level of non-gaussianity. 
First column: ghost inflation limit with flat potential. Second: ghost inflation with non-flat potential. 
Third: threshold region, higher derivative terms are as important as the tilt in the potential. Fourth: usual 
slow-roll inflation, higher derivative terms are irrelevant.} 
\end{table}

\section{Conclusions}
We have presented a new way of generating an inflationary
de Sitter phase using ghost condensation. The spectrum of density
perturbations is sharply predicted to have $n_s = 1$ up to
unobservable corrections; gravitational waves are completely unobservable. 
The non-Gaussianities are expected to be large enough to be observed.
Unfortunately because there are two relevant parameters in the
$\pi$ effective Lagrangian we can not completely fix the magnitude
of the 3-point function from the observed density perturbations; however, for all
parameters of ${\cal{O}}(1)$, the size of the effect is at the level of
the current WMAP sensitivity. Aside from the overall amplitude,
the 3-point correlator $\left\langle \frac{\delta \rho}{\rho}(k_1)
\frac{\delta \rho}{\rho}(k_2) \frac{\delta \rho}{\rho}(k_3)
\right\rangle$ as a function of the spatial momenta $k_1,k_2,k_3$ is
completely determined. 

We conclude discussing some open issues. As we stressed, all the predictions of our scenario are
independent of the details of the reheating process, analogously to standard slow-roll inflation. 
But interesting processes might occur at the end of ghost inflation. In the region where the shift 
symmetry is broken we expect that a potential for $\phi$ is generated at some level. As discussed in 
section \ref{sec:tilt}, this potential will kick $\dot\phi$ out of its stable value $\dot\phi = M^2$ and 
$\dot\phi$ will be driven back to its original value by the expansion of the universe once the shift symmetry 
is restored. The stress energy tensor of this homogeneous perturbation $\dot\pi$ redshifts like matter \cite{ghost} 
($\rho \propto a^{-3}$) so that it would be interesting to study the evolution of this fluid to see if it could 
be a viable candidate for dark matter. 

Another interesting possibility is that inflation ends with a phase transition, after which the ghost is not 
condensed anymore: Lorentz symmetry is restored. This scenario allows to raise the scale $M$, because we are not limited
anymore by the present modification of gravity. Obviously the phase transition leads us outside the regime of validity of the 
effective field theory around the ghost condensate, but it is likely that all our results about density 
perturbations still apply.

\section*{Acknowledgments}
We thank Hsin-Chia Cheng and Markus Luty for many useful discussions. We thank Leonardo Senatore for
pointing out an error in the first version of the paper. S.~M.~ is supported by the NSF
under grant PHY-0201124. M.~Z.~ is supported by NSF grants AST 0098606 and PHY 0116590 and by the David
and Lucille Packard Foundation Fellowship for Science and Engineering. N.~A.-H.~is supported by NSF under 
grant PHY-0244821 and by the David and Lucille Packard Foundation.

\appendix{Appendix: conservation of $\zeta$} \label{app:zeta}

The use of the formula (\ref{eq:shift}) for the calculation of the
gauge-invariant quantity $\zeta$ is justified in general by
the conservation of $\zeta$ during the reheating process. In this
appendix we shall explicitly check the conservation. For simplicity
we take the limit in which the time scale associated with
reheating is much shorter than the Hubble time scale and assume
that spatial components of the stress-energy tensor remain
non-singular. In this case we can use the Israel junction
condition~\cite{Israel} without a singular matter source to obtain
the matching condition for perturbations even without any
knowledge about the details of microscopic processes.

Starting with general scalar-type perturbations in the
longitudinal gauge, 
\begin{equation}
 ds^2 = -\left[1+2\Phi(t,x)\right]dt^2
 + \left[1-2\Psi(t,x)\right]a^2(t)\delta_{ij}dx^idx^j, \quad
 \phi = \phi^{(0)}(t) + \pi(t,x) \;,
\end{equation} 
we can perform an infinitesimal gauge transformation 
\begin{equation}
 \delta g_{\mu\nu} \to \delta g_{\mu\nu}
 - \xi_{\mu;\nu} - \xi_{\nu;\mu} \;,  \quad
 \pi \to \pi - \xi^{\mu}\partial_{\mu}\phi^{(0)} \;,
\end{equation} 
where
\begin{eqnarray}
 \xi_0(t,x) & = & -\frac{\pi(t_0,x)}{\dot{\phi}^{(0)}(t_0)}
  - \int_{t_0}^t\Phi(t',x)dt' \;,
  \nonumber\\
 \xi_i(t,x) & = & -a^2(t)
  \int_{t_0}^t\frac{\partial_i\xi_0(t',x)}{a^2(t')}dt' \;,
\end{eqnarray}
so that the metric and the scalar field become
\begin{eqnarray}
 ds^2 & = & -dt^2 + q_{ij}dx^idx^j\;, \nonumber\\
 q_{ij} & = &
  ( 1 -2\Psi+2H\xi_0)a^2\delta_{ij}
   -\partial_i\xi_j-\partial_j\xi_i\;, \nonumber\\
 \phi & = & \phi^{(0)}(t) + \pi(t,x) -\pi(t_0,x)
  -\dot{\phi}^{(0)}(t)\int_{t_0}^t\Phi(t',x)dt'\;.
\end{eqnarray}
Here, $H=\dot{a}/a$ and $t_0$ is the value of $t$ when the
homogeneous background $\phi^{(0)}(t)$ reaches the critical value
$\phi_0$ at which the phase transition is triggered. The
coordinate system after the gauge transformation is nothing but
the Gaussian normal coordinate system based on the reheating
hypersurface $\phi=\phi_0$. In the Gaussian normal coordinate
system, the Israel junction condition~\cite{Israel} at the
reheating hypersurface is 
\begin{equation} 
\left[q_{ij}\right] =
\left[\dot{q}_{ij}\right] = 0\;, \quad \left[X\right] \equiv
\lim_{t\to t_0+0}X(t)-\lim_{t\to t_0-0}X(t) \;. 
\end{equation} 
Thus, we obtain
\begin{equation} 
[a]=0 \; ;\qquad [H]=0 
\end{equation} 
for the unperturbed geometry, and 
\begin{equation}
\left[\frac{\pi}{\dot{\phi}^{(0)}}\right] = 0 \; ;\qquad \left[\Psi\right] = 0 
\; ;\qquad \left[\dot{\Psi}+H\Phi+\dot{H}\frac{\pi}{\dot{\phi}^{(0)}}\right] = 0 
\end{equation} 
for the linear perturbation.

Since in a pure de Sitter background the gravitational potential
at large scales decays~\cite{ghost}, it becomes vanishingly small
after a long period of de Sitter expansion. Hence, we set
$\Phi=\Psi=0$ for $t<t_0$. On the other hand, the universe after
reheating is dominated by radiation. In the radiation dominated
epoch, it is known that the Fourier components $\Phi_k$ and
$\Psi_k$ of $\Phi$ and $\Psi$, respectively, at superhorizon
scales behave as 
\begin{equation}
 \Phi_k = \Psi_k \simeq A_k + B_k\left(\frac{a(t)}{a(t_0)}\right)^{-3}
 \qquad (t>t_0) \;,
\end{equation} 
where $A_k$ and $B_k$ are constants. The coefficients $A_k$
and $B_k$ are determined by the Israel junction condition as 
\begin{equation}
 A_k = -B_k =
 \left.\frac{2}{3}H\frac{\pi_k}{\dot{\phi}^{(0)}}\right|_{t=t_0},
\end{equation} 
where $\pi_k(t)$ is the Fourier component of $\pi$. Therefore,
we obtain the following behavior of $\Phi_k$ and $\Psi_k$ after
reheating. 
\begin{equation}
\Phi_k=\Psi_k
\simeq A_k\left[1-\left(\frac{a(t_0)}{a(t)}\right)^3\right]
\to \left.\frac{2}{3}H\frac{\pi_k}{\dot{\phi}^{(0)}}\right|_{t=t_0}.
\end{equation} 
Finally, with the help of the well-known relation
$\zeta=-\frac{3}{2}\Phi$ in radiation-dominated epoch, this shows
that 
\begin{equation}
\zeta \to -\left.H\frac{\pi}{\dot{\phi}^{(0)}}\right|_{t=t_0}
\end{equation} 
and that we can safely use the formula (\ref{eq:shift}). In
the main part of this paper we shall omit the superscript $(0)$
for $\dot{\phi}^{(0)}$.


\end{document}